\documentclass[11pt]{article}
\usepackage{amsmath}

\usepackage[final]{acl}

\usepackage{times}
\usepackage{latexsym}

\usepackage[T1]{fontenc}

\usepackage[utf8]{inputenc}

\usepackage{enumitem}
\usepackage{amsfonts}
\usepackage{booktabs}

\usepackage{microtype}

\usepackage{inconsolata}

\usepackage{graphicx}

\usepackage{xcolor}
\usepackage{listings}
\usepackage{float} 

\lstdefinelanguage{json}{
    basicstyle=\ttfamily\footnotesize,
    numbers=left,
    numberstyle=\tiny,
    stepnumber=1,
    numbersep=5pt,
    showstringspaces=false,
    breaklines=true,
    frame=lines,
    morestring=[b]",
    morecomment=[l]{//},
    morecomment=[s]{/*}{*/},
    morekeywords={:}, 
    keywordstyle=\color{blue},
    stringstyle=\color{red},
    commentstyle=\color{gray}
}

\lstdefinestyle{jsonstyle}{
  basicstyle=\ttfamily\scriptsize, 
  numbers=left, numberstyle=\tiny, stepnumber=1, numbersep=6pt,
  frame=single, framerule=0.3pt, rulecolor=\color{black!25},
  breaklines=true, breakatwhitespace=true,
  columns=fullflexible, keepspaces=true, showstringspaces=false, upquote=true,
  aboveskip=6pt, belowskip=6pt, captionpos=b
}
\lstdefinestyle{promptstyle}{
  basicstyle=\ttfamily\scriptsize,
  frame=single, framerule=0.3pt, rulecolor=\color{black!25},
  breaklines=true, breakatwhitespace=true,
  columns=fullflexible, keepspaces=true, showstringspaces=false, upquote=true,
  aboveskip=6pt, belowskip=6pt, captionpos=b
}

\title{LegalSim: Multi-Agent Simulation of Legal Systems for Discovering Procedural Exploits}

\author{Sanket Badhe \\
  Rutgers University \\
  \texttt{sanketbadhe1611@gmail.com} \\}

\begin{document}
\maketitle
\begin{abstract}
We present \textsc{LegalSim}, a modular multi-agent simulation of adversarial legal proceedings that explores how AI systems can exploit procedural weaknesses in codified rules. Plaintiff and defendant agents choose from a constrained action space (for example, discovery requests, motions, meet-and-confer, sanctions) governed by a JSON rules engine, while a stochastic judge model with calibrated grant rates, cost allocations, and sanction tendencies resolves outcomes. We compare four policies: PPO, a contextual bandit with an LLM, a direct LLM policy, and a hand-crafted heuristic;  Instead of optimizing binary case outcomes, agents are trained and evaluated using effective win rate and a composite exploit score that combines opponent-cost inflation, calendar pressure, settlement pressure at low merit, and a rule-compliance margin. Across configurable regimes (e.g., bankruptcy stays, inter partes review, tax procedures) and heterogeneous judges, we observe emergent ``exploit chains'', such as cost-inflating discovery sequences and calendar-pressure tactics that remain procedurally valid yet systemically harmful. Evaluation via cross-play and Bradley-Terry ratings shows, PPO wins more often, the bandit is the most consistently competitive across opponents, the LLM trails them, and the heuristic is weakest. The results are stable in judge settings, and the simulation reveals emergent exploit chains, motivating red-teaming of legal rule systems in addition to model-level testing.
\end{abstract}

\section{Introduction}
The legal system is an adversarial process guided by dense procedural rules that shape how disputes unfold. Litigants do not only argue substance; they sequence filings, exploit timing, and impose tactical costs to influence outcomes. As AI enters legal practice, these tactics may be amplified: learning agents can search large procedural spaces, probe edge cases at scale, and coordinate strategies with speed and persistence beyond human capacity. This possibility raises questions at the intersection of natural legal language processing, multi-agent reinforcement learning, and AI safety \citep{amodei2016concreteproblemsaisafety}.

Most work in legal NLP treats models as assistive tools that classify, summarize, retrieve, or predict \citep{chalkidis-etal-2020-legal,chalkidis-etal-2022-lexglue,zhong2019jecqalegaldomainquestionanswering}. These settings assume a largely passive role for AI within human workflows. Far less is known about what happens when AI agents interact directly with codified procedure and with each other. In complex systems, agents trained to optimize rewards often uncover loopholes that remain technically compliant while socially harmful \citep{amodei2016concreteproblemsaisafety}. The legal process, with its motion practice, deadlines, and rule-based gates, is a natural domain where such behavior may emerge.

We argue that studying these dynamics requires a simulation environment that treats litigation as strategic interaction under rules. Our approach frames procedure as a structured action space with observable state, limited information, and stochastic judicial response. Agents learn over repeated play to pursue objectives that extend beyond win or loss, including cost imposition, delay, and settlement leverage under sanction risk. By varying rule sets across domains, the same environment can reveal how different procedural regimes encourage or deter exploitative behavior.

We introduce \textsc{LegalSim}, a modular multi-agent framework for adversarial legal proceedings. Plaintiff and defendant agents select structured actions validated by a JSON rule engine that encodes domain-specific procedural gates; a stochastic judge mediates outcomes via calibrated grant rates, cost allocations, and sanction tendencies. Policies include a hand-crafted heuristic baseline, a contextual bandit over tactic families, a PPO policy trained in self-play, and a direct LLM policy \citep{schulman2017proximalpolicyoptimizationalgorithms,silver2016alphago, lowe2020multiagentactorcriticmixedcooperativecompetitive,silver2017masteringchessshogiselfplay,openai2019dota2largescale,vinyals2019alphastar}. Rather than optimizing a binary case outcome, agents receive a composite exploit score that aggregates opponent-cost inflation, calendar pressure, settlement pressure conditional on low merits, and a rule-compliance margin.

\paragraph{Contributions.}
\begin{enumerate}[leftmargin=*,nosep]
\item \textbf{Formalization of litigation as a MARL environment.}
We model adversarial legal proceedings as a multi-agent environment with a structured token space, machine-readable procedural gates, and a calibrated judge model, enabling regime-agnostic studies.

\item \textbf{Discovery of emergent legal exploits.}
Through self-play, interacting agents discover strategies that were not pre-programmed, including tactics observed in practice and novel exploit chains that expose systemic vulnerabilities (cost inflation, calendar pressure, settlement leverage under sanction risk).

\item \textbf{Evaluation protocol and artifacts.}
We evaluate with head-to-head and all-against-all cross-play and fit role-symmetric Bradley-Terry-Luce ratings and robustness sweeps across judges, enabling comparison of heuristic, contextual bandit, PPO, and LLM-guided policies.

\item \textbf{AI-safety perspective on law.}
We argue for red-teaming codified legal systems themselves rather than only individual models, offering a testbed for measuring and mitigating AI-amplified procedural abuse.
\end{enumerate}

Our findings suggest an AI-safety perspective that red-teams not only models but the legal rule systems themselves. \textsc{LegalSim} offers a testbed for measuring and mitigating procedural exploitation, linking methods from legal NLP, MARL, and robustness analysis \citep{pmlr-v97-balduzzi19a,omidshafiei2019alpharankmultiagentevaluationevolution}.

\section{Background and Related Work}
Our research sits at the intersection of four areas: AI for legal reasoning, AI safety in legal contexts, multi-agent systems for strategic discovery, and the formalization of law as code. Each area is established, but their synthesis to red-team legal frameworks is new.

\subsection{AI in Legal Reasoning and Prediction}
A substantial body of Natural Legal Language Processing (NLLP) focuses on analytical and predictive tasks. Early work showed that machine learning can predict judicial outcomes from case facts \citep{aletras2016predicting}. More recent approaches leverage large language models and legal-specific pre-training, such as \textsc{Legal-BERT} \citep{chalkidis-etal-2020-legal}, achieving strong results on legal judgment prediction, document classification, and argument mining. These tools reason about or predict outcomes in a static setting; they are not designed to act as strategic agents within a procedural process. Our work shifts the focus from passive prediction to active, strategic participation.

\subsection{AI Safety and Fairness in Law}
As capabilities grow, concerns about safety and fairness have intensified. The dominant paradigm is to identify and mitigate model-level flaws, including demographic bias in predictive justice systems, as highlighted by the COMPAS investigation \citep{angwin2016machine}. Additional lines address robustness of legal text classifiers and the explainability of black-box models in service of due process \citep{bencohen2022explainability}. This work is essential, but it primarily addresses harms from systems that are wrong or biased. We study a complementary risk: harms produced by agents that are competent and strategically exploit codified rules to achieve unfair or inefficient outcomes.

\subsection{Multi-Agent Systems and Emergent Strategy}
Outside law, multi-agent reinforcement learning (MARL) has uncovered novel strategies in complex adversarial settings. Self-play has yielded superhuman policies in Go \citep{silver2016alphago} and StarCraft II \citep{vinyals2019alphastar}. Work on emergent tool use demonstrates autocurricula in competitive environments \citep{baker2020emergenttoolusemultiagent}. Classical game-theoretic analyses of litigation exist \citep{baird1994game}, but modern MARL for discovering procedural strategies from a blank slate remains underexplored. Recent agent systems that blend planning with language interaction, such as CICERO for Diplomacy \citep{meta2023cicero}, suggest feasibility for mixed-motive negotiation akin to litigation. Concurrently, agentic-risk studies examine malicious or deceptive uses of LLM agents; for example, ScamAgents demonstrates how autonomous AI agents can be architected to simulate and execute complex, human-level scam calls \citep{badhe2025scamagentsaiagentssimulate}. Our work brings these ideas into a rules-constrained, legally grounded domain.

\subsection{Computational Law and Rules-as-Code}
Formalizing legal rules in machine-readable form (\emph{rules-as-code}) is a prerequisite for procedural simulation. Foundational visions in computational law aim to represent statutes, regulations, and contracts with logical precision \citep{genesereth2005computational,surden2012computable}. Prior applications emphasize compliance checking, digital advisory tools, and expert systems. We build directly on this foundation but use the formalized ruleset as the ``physics engine'' for an adversarial simulation, enabling stress tests where intelligent agents interact strategically. While rules-as-code focuses on encoding law as written, our objective is to surface unintended and exploitative consequences that can emerge in practice.

\section{Problem Formulation}

We model litigation as a multi-agent adversarial game governed by codified procedure. Two agents, plaintiff and defendant, act in a structured environment mediated by a judge. The objective is to allow learning agents to discover strategies from the environment dynamics rather than rely on hand-coded heuristics.

\subsection{State}

At time $t$ the environment state is
\[
s_t=\{P^{\mathrm{pl}},\,P^{\mathrm{df}},\,C,\,H\},
\]
where $P^{\mathrm{pl}}$ and $P^{\mathrm{df}}$ are party states, $C$ encodes court attributes including active procedural gates and judicial tendencies, and $H$ is a structured history of filings, rulings, and citations. Party states track budgets, accumulated burden, fees, sanctions, and merits, enabling decisions conditioned on posture and history. Judicial tendencies are parameterized by a profile with a grant rate and a sanction tendency that shape probabilistic rulings.

\subsection{Actions and Gates}

At each step an agent chooses $a_t \in \mathcal{A}(s_t)$, with availability constrained by active gates:
\[
a_t \in \mathcal{A}(s_t)\quad\text{iff no gate blocks }a_t.
\]
The action space covers core procedural moves such as filing a proceeding, referencing authority, requesting discovery, moving for sanctions, changing venue, and making settlement offers. In implementation these map to structured action tokens. Gates implement rule-based blocks that delay or nullify certain actions until expiry, matching the formal constraint above. The full inventory of abstract action tokens is listed in App.~\ref{app:tokens}.

\subsection{Transition Dynamics}

A rule engine evaluates actions and updates state with deterministic and stochastic effects: impose or lift gates, allocate costs and burdens, apply sanctions, and progress the case toward termination. This supports faithful procedural interaction without prescribing strategies.

\subsection{Rewards}

Each agent’s reward combines competing litigation objectives:


\begin{align}
R_t \;=\;&\; w_1 \cdot \text{OpponentCost}_t 
          + w_2 \cdot \text{DelayCredit}_t \notag \\
        &\;+\; w_3 \cdot \text{OutcomeBonus}_t \notag \\
        &\;-\; w_4 \cdot \text{SanctionPenalty}_t
\end{align}

with weights $w_i$ tuning strategic preferences. Plaintiffs seek favorable outcomes with cost control, while defendants emphasize dismissal or delay with minimal sanction exposure. One instantiated shaping in code rewards increases in opponent burden while penalizing own cost and burden, with a terminal bonus or penalty at resolution.

\subsection{Learning Objective}

We optimize policies for both sides under discounted returns:
\[
\pi^{\mathrm{pl}},\,\pi^{\mathrm{df}}
\;=\;
\arg\max_{\pi}\;
\mathbb{E}\!\left[\sum_{t=0}^{T}\gamma^{t}\,R_t(s_t,a_t)\right],
\]
where $\gamma \in (0,1)$ is the discount factor.

\subsection{Illustrative Gate Scenario}

If a defendant files for Chapter 11\footnote{Chapter 11 of the U.S. Bankruptcy Code}, an automatic-stay gate activates and blocks motions and discovery for sixty timesteps. This induces a delay exploit for the plaintiff. After expiry, a calibrated discovery sequence can inflate the defendant’s costs if judicial sanctions are unlikely, yielding a cost-inflation exploit while each step remains procedurally valid.

This formulation specifies the state, action, gating, transition, reward, and objective needed to study emergent procedural exploits within a principled multi-agent reinforcement learning setting.

\section{System Architecture} \label{sec:system-architecture}

Our system's architecture is organized into a three-tiered structure: the \textsc{LegalSim} environment, the agents that interact within it, and a training and analysis harness. This design separates the core simulation logic from the agent policies and experiment orchestration, which enables modularity and the easy substitution of components without changing core interfaces.

\subsection{Environment Layer}

The environment layer is a domain-agnostic litigation simulator. It is driven by a rules-as-code engine that processes an abstract action space and updates the simulation state based on a set of JSON rules. Each environment instance maintains the state of two adversarial parties, tracking their budgets, accumulated burdens, and merits. It also incorporates a stochastic judge profile that influences probabilistic rulings, such as the granting of motions or the imposition of sanctions.

\subsubsection{Rules-as-code Engine}

The core of the environment is a JSON rule engine (a finite set of state–action predicates with effect handlers) that loads procedural rules and gates (temporary action blocks) from JSON files \citep{governatori2015norms}. Rules are defined by when conditions (e.g., a specific action is taken) and effects that modify the environment state. Effects can include applying costs, transferring burdens, and, most importantly, activating procedural gates. The RulesOracle component provides an approximation of legal principles like proportionality and sanction risk, which are parameterized to allow for policy studies rather than strict encoding of legal doctrine. We include rule example at App.~\ref{app:rules}.

\subsubsection{Procedural regimes.}
The environment swaps procedural regimes by loading different JSON rule files without code changes.
In our experiments we use a default bankruptcy regime and also include domain-specific sets for patent, tax collection, immigration, and corporate disputes.
Each regime defines gates (temporary blocks on actions) and effects that shape costs, burdens, delay, and sanctions.

\subsubsection{Action Interface and State}
Agents interact with the environment by emitting abstract, tokenized actions. In total, there are 13 possible action tokens, including representative examples such as \texttt{REQUEST\_DOCS}, \texttt{FILE\_PROCEEDING}, and \texttt{SETTLEMENT\_OFFER}. The environment validates these tokens against a predefined schema, ensuring the action space remains structured and the simulation runs smoothly. The environment state includes party-specific metrics (budget, burden, merits) and global context from the judge, such as grant rate and sanction tendency.

\subsection{Agent Layer}

The agent layer supports multiple policy families that are swappable through a common interface. This allows for a mix-and-match approach in experiments, where different agent types can be pitted against each other. The policies include:

\begin{enumerate}[leftmargin=*,nosep]
\item \textbf{Heuristic Policy:} A hand-coded, rule-based baseline that makes decisions based on simple, pre-defined logic related to costs and burdens.
\item \textbf{LLM-driven Policy:} A policy that queries a large language model (LLM) for a reasoned action. It uses few-shot prompting and enforces a strict JSON output contract to ensure the LLM's free-form reasoning can be translated into a valid action token. Throughout this paper, all LLM calls use OpenAI’s GPT-4o. 
\item \textbf{Contextual Bandit Policy:} A hybrid policy that first uses a contextual bandit to select a high-level "tactic" (e.g., DELAY, BURDEN\_OPP), and then uses the LLM to propose a specific action consistent with that tactic. \citep{Li_2010}
\item \textbf{PPO Policy:} A policy based on Proximal Policy Optimization (PPO), a reinforcement learning algorithm that learns to select actions from the environment's observations \citep{schulman2017proximalpolicyoptimizationalgorithms,silver2016alphago}.
\end{enumerate}

\subsection{Training and Evaluation Harness}

The harness coordinates self-play experiments, enforcing role alternation and judge rotation, scheduling learning updates for PPO and the contextual bandit, and validating all emitted action tokens.

\subsubsection{Episode Flow}

A single episode unfolds as follows: the harness initializes the environment, agents observe the state and propose actions, the environment validates and executes these actions, and the state advances. This process repeats until a termination condition is met (e.g., budget exhaustion, settlement, or maximum steps). At termination, composite exploit metrics are calculated, and learning updates are applied to the agents' policies if enabled.

\section{Experiments and Evaluation}

We evaluate \textsc{LegalSim} under a controlled protocol that alternates roles each game, sweeps ten random seeds, and rotates between two judge profiles: \emph{permissive} (grant\_rate 0.65, sanction\_tendency 0.25, calendar\_load 0.55) and \emph{strict} (0.35, 0.70, 0.60). Domains are loaded from JSON rule files; unless noted, we use the default regime. 

\paragraph{Policies and training.}
We evaluate the four policy families introduced in Sec.~\ref{sec:system-architecture} (Heuristic, LLM, Contextual Bandit, PPO). The \emph{Heuristic} is non-learning. The \emph{LLM} policy uses a single API model at inference time and emits JSON-constrained tokens without any fine-tuning. The \emph{Contextual Bandit} selects a high-level tactic via an $\epsilon$-greedy linear contextual bandit with a bias term ($\epsilon{=}0.1$, learning rate $0.05$), then asks the same LLM to instantiate a concrete token; it performs one SGD update per episode on the terminal composite reward. The \emph{PPO} agent is an actor critic over the discrete token set with a 13-D observation (budgets, burdens, judge features, merits, progress, gate summaries); both actor and critic are two-layer MLPs (64 Tanh units each) trained with Adam ($3\times10^{-4}$), $\gamma{=}0.99$, GAE $\lambda{=}0.95$, clip $\epsilon{=}0.2$, and entropy coefficient $0.005$. PPO optimizes the shaped reward

\begin{multline}
r_t \;=\; 0.20\,\Delta \text{(opponent burden)} 
          - 0.01\,\Delta \text{(own cost)} \\
          - 0.01\,\Delta \text{(own burden)}\,.
\end{multline}

with a terminal bonus of $+5$ (plaintiff win) and $-5$ (defendant win), is trained for 300 episodes against the Heuristic while alternating judges by episode, and is then frozen for evaluation.

\paragraph{Environment and rules.}
Litigation is modeled as a turn-based process with a rules-as-code core. Agents emit tokens such as \texttt{REQUEST\_DOCS}, \texttt{FILE\_MOTION}, \texttt{MOVE\_SANCTIONS}, \texttt{MEET\_CONFER}, \texttt{SETTLEMENT\_OFFER}, and \texttt{FILE\_PROCEEDING}. A JSON rule engine maps state--action conditions to cost transfers, burden updates, temporary gates that block actions, and judge-sensitive sanction events; a \texttt{RulesOracle} provides proportionality and sanction-risk proxies. Each episode tracks budgets, burdens, merits, fees, sanctions, active gates, and the judge profile.

\paragraph{Protocols.}
We use two complementary designs. (i) \emph{Head-to-head}: selected pairs play under both judges with alternating roles; we log token sequences, rulings, and per-role metrics. (ii) \emph{Cross-play league}: all policies play all others across seeds and judges. From these games we build a role-symmetric payoff matrix $A$ whose entry
\begin{align}
A_{ij} \;=\; \mathbb{E}\!\big[ 
   & \text{(plaintiff composite)} \notag \\
   & - \text{(defendant composite)} 
\big]
\end{align}
when policy $i$ faces $j$, flipping the sign when roles swap so cells are comparable. We also fit role-symmetric Bradley-Terry-Luce (BTL) ratings with a sum-to-zero constraint \citep{bradley1952rank}:
\[
\Pr(i \succ j) \;=\; \sigma(s_i - s_j)\,,\qquad \sigma(x)=\frac{1}{1+e^{-x}},
\]
and report 95\% bootstrap confidence intervals over 500 resamples.
Prompt templates used during evaluation are reproduced in App.~\ref{app:prompts}

\begin{figure}[t]
  \centering
  \includegraphics[width=\columnwidth]{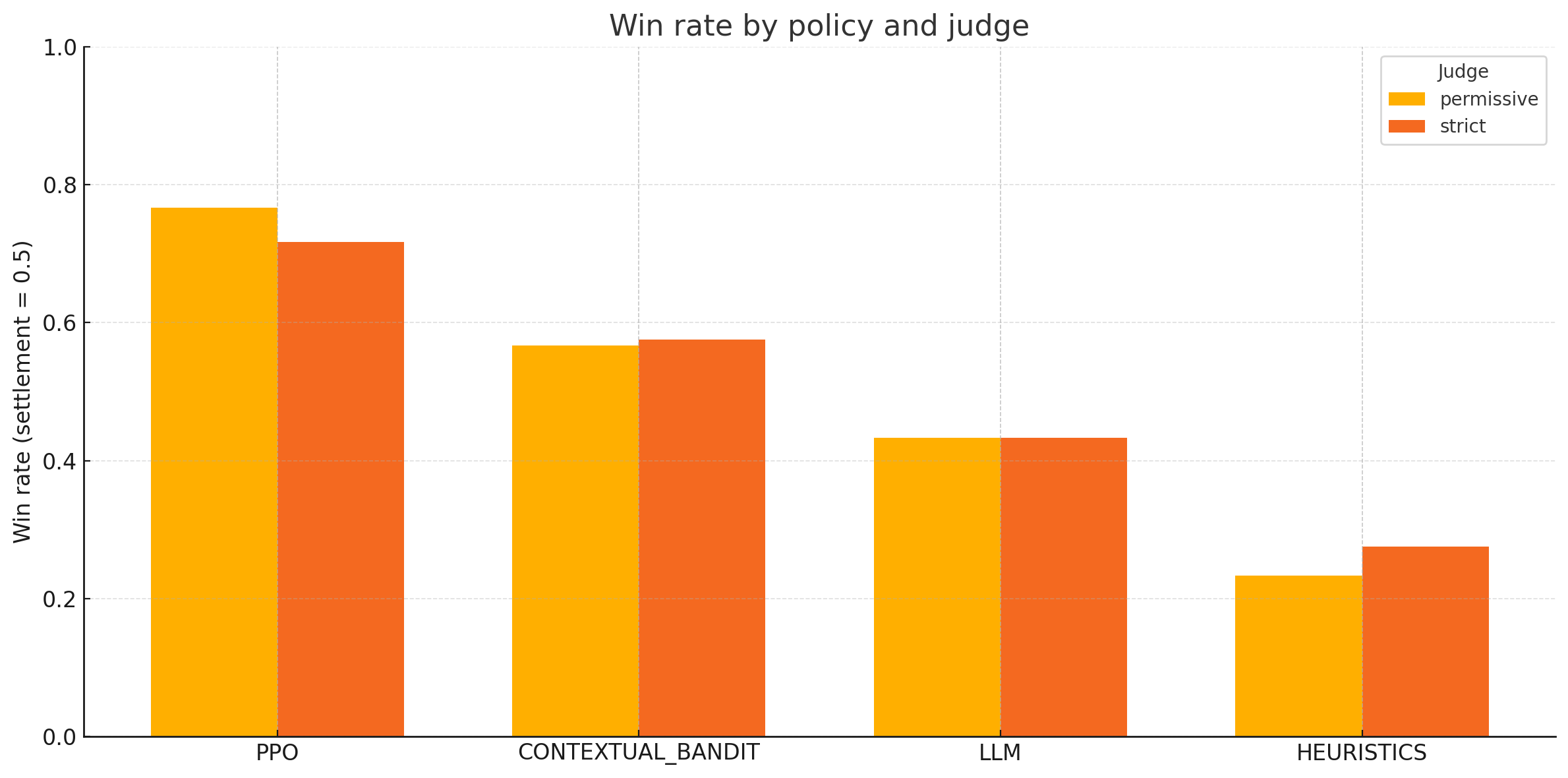}
  \caption{Win rate by policy \& judge (settlement $=0.5$). Bars show mean effective win rate across ten seeds under \emph{permissive} and \emph{strict} judges; higher is better for the policy.}
  \label{fig:winrate}
\end{figure}

\paragraph{Overall effectiveness.}
The win-rate analysis in Figure~\ref{fig:winrate} (Win-rate by policy \& judge) shows a consistent ordering: PPO attains the highest effective win rate overall, followed by the contextual bandit, then the LLM policy, with the heuristic trailing. The same ordering holds within each judge profile. This indicates that learning a direct policy over the token space (PPO) converts the observable state into match wins more reliably than tactic selection with LLM instantiation (contextual bandit) or purely generative action proposals (LLM). Table~\ref{tab:eval-summary} quantifies these differences alongside mean composite exploit scores for plaintiff/defendant roles.

\begin{table*}[t]
\centering
\small
\setlength{\tabcolsep}{6pt}
\begin{tabular}{lrrrrrrr}
\toprule
Policy & Win rate$_\mathrm{eff}$ & Flag rate & $\overline{C}_\mathrm{pl}$ & $\overline{C}_\mathrm{df}$ & BTL & BTL CI$_\mathrm{low}$ & BTL CI$_\mathrm{high}$ \\
\midrule
contextual\_bandit & 0.571 & 0.958 & 1.130 & 0.948 & 99.6 & 74.3 & 124.7 \\
llm                & 0.433 & 1.000 & 15.789 & 5.509 & 95.9 & 70.3 & 120.9 \\
ppo                & 0.742 & 0.958 & 1.280 & 0.845 & -97.6 & -123.0 & -72.6 \\
heuristic          & 0.254 & 1.000 & 39.995 & 8.337 & -97.9 & -122.6 & -71.6 \\
\bottomrule
\end{tabular}
\caption{Evaluation summary by policy. Win rate$_\mathrm{eff}$ treats settlements as 0.5. $\overline{C}_\mathrm{pl}$ and $\overline{C}_\mathrm{df}$ are mean composite exploit scores for plaintiff/defendant roles. BTL and 95\% bootstrap CIs are from role-symmetric Bradley-Terry-Luce fits on the cross-play league.}
\label{tab:eval-summary}
\end{table*}

\begin{figure*}[t]
  \centering
  \includegraphics[width=\textwidth]{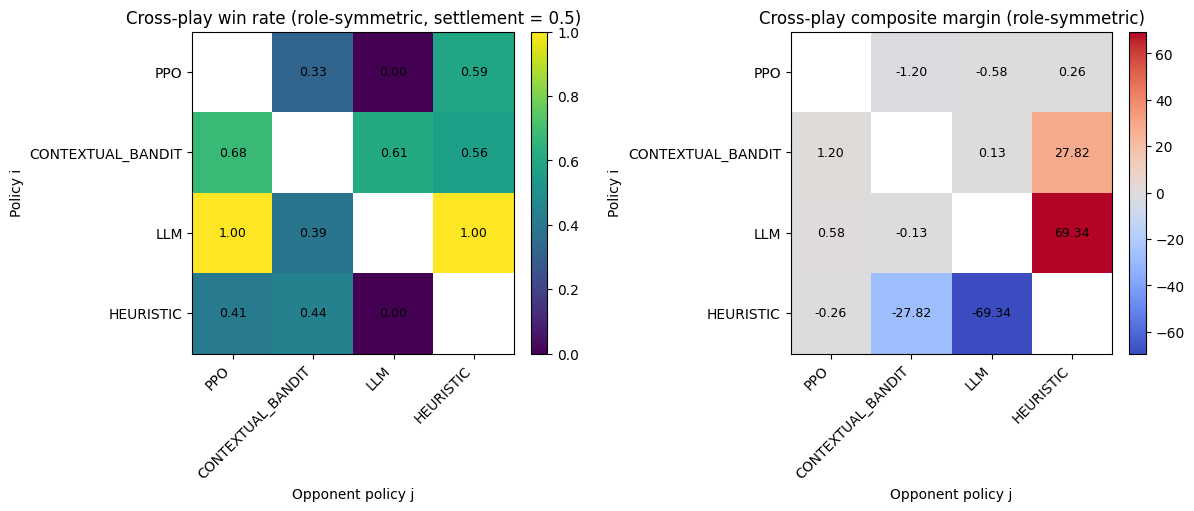}
  \caption{\textbf{Cross-play performance heatmaps (role-symmetric).} 
  Left: win rate with settlements counted as $0.5$, averaged over both role assignments. 
  Right: composite margin (plaintiff composite $-$ defendant composite) with the sign flipped when roles swap. 
  Rows index the \emph{row} policy $i$ and columns the \emph{opponent} policy $j$; numbers are cell means across seeds and judges. 
  Higher (warmer) values indicate that the row policy systematically outperforms (or exerts more procedural pressure than) the column policy.}
  \label{fig:crossplay}
\end{figure*}

\paragraph{Who beats whom (meta-game structure).}
Figure~\ref{fig:crossplay} summarizes pairwise performance using two role-symmetric metrics: (i) win rate with settlements counted as $0.5$, averaged across both role assignments for each policy pair (row policy $i$ vs.\ column policy $j$); and (ii) composite margin, the mean difference in exploit score \((\text{plaintiff composite} - \text{defendant composite})\) with the sign flipped when roles swap, so positive values indicate that the row policy systematically exerts more procedural pressure than the column policy. The heatmaps yield a consistent ordering: the Contextual Bandit dominates win rates $\ge 0.56$ against all opponents and positive margins (largest vs.\ Heuristic, modest vs.\ LLM); the LLM policy is second, clearly ahead of PPO and Heuristic; PPO shows advantage only over the Heuristic; and the Heuristic is uniformly weakest. This ranking holds in both outcome space (win rate) and pressure space (composite margin), indicating the bandit’s broad competitiveness across the meta-game.
An example episode underlying a high-margin cell is unpacked in App.~\ref{app:exploit}.

\paragraph{Exploitiveness metrics.}
Each episode produces per-role components already defined in the environment: (i) \emph{opponent cost inflation} (opponent fees divided by own fees), (ii) \emph{calendar pressure} (opponent burden divided by $1+$ own burden), (iii) \emph{settlement pressure at low merit} (settlement offers times $1-$ own merits, clipped), and (iv) \emph{rule-compliance margin} (a penalty for self-sanctions). The composite exploit score is a fixed weighted sum of these components (0.35/0.25/0.25/0.15). We summarize these per policy with means and standard errors and also report the \emph{flag rate}, the fraction of episodes with composite $\ge 0.6$.

\paragraph{Exploitiveness results.}
Applying these definitions, Table~\ref{tab:exploitiveness-summary} reports per–policy $\times$ judge means, standard errors, flag rates, and episode counts. Two patterns emerge. First, the Heuristic and LLM policies produce very high composite scores with near-ubiquitous flagging across judges, indicating heavy procedural pressure. Second, PPO and the Contextual Bandit maintain composites near~1 with non-maximal flag rates, and show judge sensitivity (the bandit declines under the strict judge, whereas PPO ticks up slightly). Together with Figure~\ref{fig:winrate}, this confirms that effectiveness (win rate) and exploitiveness are related but not identical: PPO converts state to wins while applying less extreme procedural pressure than the Heuristic or LLM, and the bandit sits between these extremes.

\begin{table*}[t]
\centering
\small
\setlength{\tabcolsep}{6pt}
\begin{tabular}{llrrrr}
\toprule
\textbf{Policy} & \textbf{Judge} & \textbf{Mean Composite} & \textbf{SE Composite} & \textbf{Flag Rate} & \textbf{N Episodes} \\
\midrule
Contextual Bandit & Permissive & 1.18 & 0.145 & 0.83 & 60 \\
Contextual Bandit & Strict     & 0.90 & 0.098 & 0.70 & 60 \\
Heuristic         & Permissive & 23.24 & 5.353 & 1.00 & 60 \\
Heuristic         & Strict     & 25.09 & 5.893 & 0.98 & 60 \\
LLM               & Permissive & 10.92 & 2.558 & 1.00 & 60 \\
LLM               & Strict     & 10.38 & 2.650 & 1.00 & 60 \\
PPO               & Permissive & 1.01 & 0.054 & 0.78 & 60 \\
PPO               & Strict     & 1.12 & 0.065 & 0.83 & 60 \\
\bottomrule
\end{tabular}
\caption{Exploitiveness summary by policy and judge. \emph{Mean Composite} is the average composite exploit score $C$; \emph{SE Composite} is the standard error; \emph{Flag Rate} is the fraction of episodes with $C \ge 0.6$; and \emph{N Episodes} is the number of episodes summarized.}
\label{tab:exploitiveness-summary}
\end{table*}

\paragraph{Judge effects.}
Breaking out the bars in Figure~\ref{fig:winrate} by judge shows that absolute win rates shift with judicial temperament, but the relative ordering of policies remains stable. On the permissive judge, motion-driven strategies benefit more; on the strict judge, margins compress but the ranking persists. This mirrors the sanction and grant-rate sensitivities in Table~\ref{tab:eval-summary} and the stratified means in Table~\ref{tab:robustness}.

\paragraph{Robustness.}
We stress-test the policies in two simple ways: (i) we make the judge more likely to impose sanctions, and (ii) we add random \(\pm 10\%\)–\(20\%\) perturbations to cost and burden parameters to mimic modeling noise. For each policy and judge profile we then recompute two summaries, the mean composite exploit score and the fraction of episodes that are flagged (\(C \ge 0.6\)), and present them as a policy-by-stress matrix (Table~\ref{tab:robustness}).

Across all stress tests, the qualitative ordering of policies does not change: PPO remains strongest on outcomes, the contextual bandit is generally second, the LLM trails, and the heuristic is consistently weakest. Making sanctions stricter reliably lowers exploit scores and flag rates, with the biggest reductions for strategies that lean on filing volume and burden (LLM, then bandit), while PPO is least affected. Injecting cost/burden noise increases variability but does not reverse pairwise rankings. In short, the effects we report are stable to reasonable procedural and parameter changes; stricter sanction regimes act as a partial brake on exploit-heavy behavior without reshuffling the policy hierarchy.

\begin{table}[t]
\centering
\small
\setlength{\tabcolsep}{6pt}
\begin{tabular}{lrrr}
\toprule
\multicolumn{4}{c}{\textbf{Sanction tendency sweep}}\\
\midrule
Sanction & Mean $C$ & 95\% low & 95\% high \\
\midrule
0.10 & 24.63 & 16.61 & 32.52 \\
0.25 & 24.62 & 16.84 & 32.68 \\
0.50 & 24.19 & 16.86 & 32.04 \\
0.75 & 24.44 & 17.40 & 32.07 \\
0.90 & 25.69 & 17.54 & 33.91 \\
\midrule
\multicolumn{4}{c}{\textbf{Parameter noise sweep}}\\
\midrule
Noise & Mean $C$ & 95\% low & 95\% high \\
\midrule
$-0.20$ & 32.22 & 18.54 & 46.43 \\
$-0.10$ & 27.15 & 15.56 & 39.12 \\
$0.00$  & 26.49 & 15.20 & 38.26 \\
$+0.10$ & 24.48 & 13.85 & 35.26 \\
$+0.20$ & 22.02 & 12.31 & 32.01 \\
\bottomrule
\end{tabular}
\caption{Robustness summary (aggregate). Mean composite exploit score $C$ and bootstrap 95\% CIs under (top) sanction-tendency sweep and (bottom) multiplicative parameter-noise sweep applied to costs/burdens. Values are aggregated across policies and both judges, with $n_{\text{episodes}} = 60$ simulation runs per configuration.}
\label{tab:robustness}
\end{table}

\paragraph{Reconciling win rate with BTL ratings.}
 BTL summarizes global competitiveness from the full cross-play, not just wins. It can rank a policy higher when it draws fewer severe losses and plays most opponents close, even if its raw win rate is slightly lower. In our run, PPO tops Table~\ref{tab:exploitiveness-summary} for win rate; BTL places the contextual bandit and LLM closer in the middle of the meta-game (with overlapping confidence intervals), while the heuristic sits clearly below. This is consistent with Figure~\ref{fig:crossplay} showing near-zero margins between the mid-tier policies and large negative margins concentrated in the heuristic row and column.

\section{Defense, and Mitigation}

\subsection*{Risks}
Optimizing agents can find strategies that are legal but harmful, turning procedural gaps into cost and delay weapons \citep{amodei2016concreteproblemsaisafety}. Law is especially exposed because it is highly codified, adversarial, and variably adjudicated. This risk of ``reward hacking,'' where an agent satisfies the literal specification of a reward function in an unintended way, is a fundamental challenge in aligning AI with complex, real-world objectives \citep{leike2018scalableagentalignmentreward}.

\paragraph{Beyond accuracy.}
The problem is not only wrong predictions but system-level exploits that emerge when agents play the rules. League and cross-play results echo findings in open multi-agent games: small tactical gains can snowball into undesirable equilibria \citep{pmlr-v97-balduzzi19a}. Evaluating performance in such ecologies requires methods beyond simple win-rates, such as AlphaRank or Bradley-Terry models, to capture the full matrix of strategic interactions \citep{omidshafiei2019alpharankmultiagentevaluationevolution,bradley1952rank}.

\paragraph{Design-time defenses.}
Harden procedures before deployment: introduce light randomization in scheduling, add rule linting to detect long burden-inflating chains, and tie cost shifting to burden ratios so exploit-heavy sequences become expensive.

\paragraph{Governance.}
Require pre-deployment red-teaming in a rules-as-code sandbox, disclosure of agent capabilities, and auditable reports with cross-play matrices, BTL ratings, and exploit dashboards. This aligns with emerging AI governance standards, such as the NIST AI Risk Management Framework, which emphasizes continuous testing, evaluation, and risk mitigation throughout the AI lifecycle 

The core risk is not that agents can win, but that they can steer rule-driven systems toward exploit-heavy equilibria. Simulation-first analysis, league benchmarks, and targeted procedural guardrails provide a practical path to measure and mitigate these effects.

\section{Ethical Considerations}
This work examines how agentic AI might exploit codified procedure with the goal of improving safety and fairness in legal automation by identifying failure modes before real-world use. By surfacing and quantifying “exploit chains,” we aim to support due-process values and reduce risks. We acknowledge dual-use concerns: the same insights could enable misuse. To mitigate this, we confine our analysis to simulation, avoid jurisdiction-specific guidance, and emphasize safeguards. \textsc{LegalSim} is a research simulator and not legal advice.

\section{Limitations}
Our simulator necessarily abstracts complex laws and institutional practice: rules are encoded at a coarse level, the judge model is parametric and stationary, and strategy discovery is constrained by a tokenized action space. The exploit metrics and their weights, though motivated, are ultimately design choices that may not capture the full spectrum of welfare-relevant harms. Similarly, policy coverage remains narrow (four families) and training horizons modest, such that stronger or more sample-efficient methods could shift the observed rankings. These results should therefore not be assumed to generalize across jurisdictions, case types, or institutional settings, especially since the environment omits strategic behavior by non-party actors such as regulators or multi-judge panels.

A central limitation is that our findings have not yet been grounded in real-world judicial data or case law. While the experiments reveal how artificial agents may exploit procedural rules in silico, we have not examined whether comparable exploitative dynamics occur in practice, nor how judicial actors (e.g., judges, clerks, regulators) adapt to mitigate such behaviors.

Finally, we emphasize that simulations cannot substitute for legal or ethical judgment. Insights derived here should inform, but never replace, human governance and procedural safeguards.

\section{Conclusion}
We introduced \textsc{LegalSim}, a modular multi-agent simulation that treats procedure as rules-as-code and measures how AI-driven strategies can exploit legal process. Across head-to-head and cross-play evaluations, we observed consistent ordering among policies, documented emergent “exploit chains,” and quantified exploitiveness with outcome and pressure centric metrics. These results frame procedural robustness as an AI-safety problem: not only how models behave, but how codified rules can be gamed. The framework provides a controlled setting to study defenses, e.g. randomized gates, human review for high-impact actions, and system-level red-teaming, before deployment in real practice. We hope this work catalyzes collaboration between NLP, MARL, and legal communities on measuring and mitigating AI-amplified procedural abuse.


\begin{thebibliography}{25}
\providecommand{\natexlab}[1]{#1}

\bibitem[{Aletras et~al.(2016)Aletras, Tsarapatsanis, Preoţiuc-Pietro, and Lampos}]{aletras2016predicting}
Nikolaos Aletras, Dimitrios Tsarapatsanis, Daniel Preoţiuc-Pietro, and Vasileios Lampos. 2016.
\newblock \href {https://doi.org/10.7717/peerj-cs.93} {Predicting judicial decisions of the european court of human rights: a natural language processing perspective}.
\newblock \emph{PeerJ Computer Science}, 2:e93.

\bibitem[{Amodei et~al.(2016)Amodei, Olah, Steinhardt, Christiano, Schulman, and Mané}]{amodei2016concreteproblemsaisafety}
Dario Amodei, Chris Olah, Jacob Steinhardt, Paul Christiano, John Schulman, and Dan Mané. 2016.
\newblock \href {https://arxiv.org/abs/1606.06565} {Concrete problems in ai safety}.
\newblock \emph{Preprint}, arXiv:1606.06565.

\bibitem[{Angwin et~al.(2016)Angwin, Larson, Mattu, and Kirchner}]{angwin2016machine}
Julia Angwin, Jeff Larson, Surya Mattu, and Lauren Kirchner. 2016.
\newblock Machine bias: There's software used across the country to predict future criminals. and it's biased against blacks.
\newblock \url{https://www.propublica.org/article/machine-bias-risk-assessments-in-criminal-sentencing}.

\bibitem[{Badhe(2025)}]{badhe2025scamagentsaiagentssimulate}
Sanket Badhe. 2025.
\newblock \href {https://arxiv.org/abs/2508.06457} {Scamagents: How ai agents can simulate human-level scam calls}.
\newblock \emph{Preprint}, arXiv:2508.06457.

\bibitem[{Baird et~al.(1994)Baird, Gertner, and Picker}]{baird1994game}
Douglas~G Baird, Robert~H Gertner, and Randal~C Picker. 1994.
\newblock \emph{Game theory and the law}.
\newblock Harvard University Press.

\bibitem[{Baker et~al.(2020)Baker, Kanitscheider, Markov, Wu, Powell, McGrew, and Mordatch}]{baker2020emergenttoolusemultiagent}
Bowen Baker, Ingmar Kanitscheider, Todor Markov, Yi~Wu, Glenn Powell, Bob McGrew, and Igor Mordatch. 2020.
\newblock \href {https://arxiv.org/abs/1909.07528} {Emergent tool use from multi-agent autocurricula}.
\newblock \emph{Preprint}, arXiv:1909.07528.

\bibitem[{Balduzzi et~al.(2019)Balduzzi, Garnelo, Bachrach, Czarnecki, Perolat, Jaderberg, and Graepel}]{pmlr-v97-balduzzi19a}
David Balduzzi, Marta Garnelo, Yoram Bachrach, Wojciech Czarnecki, Julien Perolat, Max Jaderberg, and Thore Graepel. 2019.
\newblock \href {https://proceedings.mlr.press/v97/balduzzi19a.html} {Open-ended learning in symmetric zero-sum games}.
\newblock In \emph{Proceedings of the 36th International Conference on Machine Learning}, volume~97 of \emph{Proceedings of Machine Learning Research}, pages 434--443. PMLR.

\bibitem[{Bradley and Terry(1952)}]{bradley1952rank}
Ralph~Allan Bradley and Milton~E. Terry. 1952.
\newblock \href {https://doi.org/10.2307/2334029} {Rank analysis of incomplete block designs: I. the method of paired comparisons}.
\newblock \emph{Biometrika}, 39(3/4):324--345.

\bibitem[{Chalkidis et~al.(2020)Chalkidis, Fergadiotis, Malakasiotis, Aletras, and Androutsopoulos}]{chalkidis-etal-2020-legal}
Ilias Chalkidis, Manos Fergadiotis, Prodromos Malakasiotis, Nikolaos Aletras, and Ion Androutsopoulos. 2020.
\newblock \href {https://doi.org/10.18653/v1/2020.findings-emnlp.261} {{LEGAL}-{BERT}: The muppets straight out of law school}.
\newblock In \emph{Findings of the Association for Computational Linguistics: EMNLP 2020}, pages 2898--2904, Online. Association for Computational Linguistics.

\bibitem[{Chalkidis et~al.(2022)Chalkidis, Jana, Hartung, Bommarito, Androutsopoulos, Katz, and Aletras}]{chalkidis-etal-2022-lexglue}
Ilias Chalkidis, Abhik Jana, Dirk Hartung, Michael Bommarito, Ion Androutsopoulos, Daniel Katz, and Nikolaos Aletras. 2022.
\newblock \href {https://doi.org/10.18653/v1/2022.acl-long.297} {{L}ex{GLUE}: A benchmark dataset for legal language understanding in {E}nglish}.
\newblock In \emph{Proceedings of the 60th Annual Meeting of the Association for Computational Linguistics (Volume 1: Long Papers)}, pages 4310--4330, Dublin, Ireland. Association for Computational Linguistics.

\bibitem[{FAIR et~al.(2022)FAIR, Bakhtin, Brown, Dinan, Farina, Flaherty, Fried, Goff, Gray, Hu, Jacob, Komeili, Konath, Kwon, Lerer, Lewis, Miller, Mitts, Renduchintala, Roller, Rowe, Shi, Spisak, Wei, Wu, Zhang, and Zijlstra}]{meta2023cicero}
FAIR, Anton Bakhtin, Noam Brown, Emily Dinan, Gabriele Farina, Colin Flaherty, Daniel Fried, Andrew Goff, Jonathan Gray, Hengyuan Hu, Athul~Paul Jacob, Mojtaba Komeili, Karthik Konath, Minae Kwon, Adam Lerer, Mike Lewis, Alexander~H. Miller, Sasha Mitts, Adithya Renduchintala, and 8 others. 2022.
\newblock \href {https://doi.org/10.1126/science.ade9097} {Human-level play in the game of <i>diplomacy</i> by combining language models with strategic reasoning}.
\newblock \emph{Science}, 378(6624):1067--1074.

\bibitem[{Genesereth(2005)}]{genesereth2005computational}
Michael Genesereth. 2005.
\newblock \href {https://doi.org/10.1145/1165485.1165517} {Computational law}.
\newblock In \emph{Proceedings of the 10th International Conference on Artificial Intelligence and Law (ICAIL '05)}, pages 12--13, New York, NY, USA. ACM.

\bibitem[{Governatori et~al.(2011)Governatori, Olivieri, Scannapieco, and Cristani}]{governatori2015norms}
Guido Governatori, Francesco Olivieri, Simone Scannapieco, and Matteo Cristani. 2011.
\newblock Designing for compliance: Norms and goals.
\newblock In \emph{Rule-Based Modeling and Computing on the Semantic Web}, pages 282--297, Berlin, Heidelberg. Springer Berlin Heidelberg.

\bibitem[{Leike et~al.(2018)Leike, Krueger, Everitt, Martic, Maini, and Legg}]{leike2018scalableagentalignmentreward}
Jan Leike, David Krueger, Tom Everitt, Miljan Martic, Vishal Maini, and Shane Legg. 2018.
\newblock \href {https://arxiv.org/abs/1811.07871} {Scalable agent alignment via reward modeling: a research direction}.
\newblock \emph{Preprint}, arXiv:1811.07871.

\bibitem[{Li et~al.(2010)Li, Chu, Langford, and Schapire}]{Li_2010}
Lihong Li, Wei Chu, John Langford, and Robert~E. Schapire. 2010.
\newblock \href {https://doi.org/10.1145/1772690.1772758} {A contextual-bandit approach to personalized news article recommendation}.
\newblock In \emph{Proceedings of the 19th international conference on World wide web}, WWW ’10, page 661–670. ACM.

\bibitem[{Lowe et~al.(2020)Lowe, Wu, Tamar, Harb, Abbeel, and Mordatch}]{lowe2020multiagentactorcriticmixedcooperativecompetitive}
Ryan Lowe, Yi~Wu, Aviv Tamar, Jean Harb, Pieter Abbeel, and Igor Mordatch. 2020.
\newblock \href {https://arxiv.org/abs/1706.02275} {Multi-agent actor-critic for mixed cooperative-competitive environments}.
\newblock \emph{Preprint}, arXiv:1706.02275.

\bibitem[{Omidshafiei et~al.(2019)Omidshafiei, Papadimitriou, Piliouras, Tuyls, Rowland, Lespiau, Czarnecki, Lanctot, Perolat, and Munos}]{omidshafiei2019alpharankmultiagentevaluationevolution}
Shayegan Omidshafiei, Christos Papadimitriou, Georgios Piliouras, Karl Tuyls, Mark Rowland, Jean-Baptiste Lespiau, Wojciech~M. Czarnecki, Marc Lanctot, Julien Perolat, and Remi Munos. 2019.
\newblock \href {https://arxiv.org/abs/1903.01373} {$\alpha$-rank: Multi-agent evaluation by evolution}.
\newblock \emph{Preprint}, arXiv:1903.01373.

\bibitem[{OpenAI et~al.(2019)OpenAI, :, Berner, Brockman, Chan, Cheung, Dębiak, Dennison, Farhi, Fischer, Hashme, Hesse, Józefowicz, Gray, Olsson, Pachocki, Petrov, d.~O.~Pinto, Raiman, Salimans, Schlatter, Schneider, Sidor, Sutskever, Tang, Wolski, and Zhang}]{openai2019dota2largescale}
OpenAI, :, Christopher Berner, Greg Brockman, Brooke Chan, Vicki Cheung, Przemysław Dębiak, Christy Dennison, David Farhi, Quirin Fischer, Shariq Hashme, Chris Hesse, Rafal Józefowicz, Scott Gray, Catherine Olsson, Jakub Pachocki, Michael Petrov, Henrique~P. d.~O.~Pinto, Jonathan Raiman, and 8 others. 2019.
\newblock \href {https://arxiv.org/abs/1912.06680} {Dota 2 with large scale deep reinforcement learning}.
\newblock \emph{Preprint}, arXiv:1912.06680.

\bibitem[{Richmond et~al.(2023)Richmond, Muddamsetty, Gammeltoft-Hansen, Olsen, and Moeslund}]{bencohen2022explainability}
Karen Richmond, Satya Muddamsetty, Thomas Gammeltoft-Hansen, Henrik Olsen, and Thomas Moeslund. 2023.
\newblock \href {https://doi.org/10.1007/s44206-023-00081-z} {Explainable ai and law: An evidential survey}.
\newblock \emph{Digital Society}, 3.

\bibitem[{Schulman et~al.(2017)Schulman, Wolski, Dhariwal, Radford, and Klimov}]{schulman2017proximalpolicyoptimizationalgorithms}
John Schulman, Filip Wolski, Prafulla Dhariwal, Alec Radford, and Oleg Klimov. 2017.
\newblock \href {https://arxiv.org/abs/1707.06347} {Proximal policy optimization algorithms}.
\newblock \emph{Preprint}, arXiv:1707.06347.

\bibitem[{Silver et~al.(2016)Silver, Huang, Maddison, Guez, Sifre, van~den Driessche, Schrittwieser, Antonoglou, Panneershelvam, Lanctot, Dieleman, Grewe, Nham, Kalchbrenner, Sutskever, Lillicrap, Leach, Kavukcuoglu, Graepel, and Hassabis}]{silver2016alphago}
David Silver, Aja Huang, Christopher~J. Maddison, Arthur Guez, Laurent Sifre, George van~den Driessche, Julian Schrittwieser, Ioannis Antonoglou, Veda Panneershelvam, Marc Lanctot, Sander Dieleman, Dominik Grewe, John Nham, Nal Kalchbrenner, Ilya Sutskever, Timothy Lillicrap, Madeleine Leach, Koray Kavukcuoglu, Thore Graepel, and Demis Hassabis. 2016.
\newblock \href {http://www.nature.com/nature/journal/v529/n7587/full/nature16961.html} {Mastering the game of go with deep neural networks and tree search}.
\newblock \emph{Nature}, 529:484--503.

\bibitem[{Silver et~al.(2017)Silver, Hubert, Schrittwieser, Antonoglou, Lai, Guez, Lanctot, Sifre, Kumaran, Graepel, Lillicrap, Simonyan, and Hassabis}]{silver2017masteringchessshogiselfplay}
David Silver, Thomas Hubert, Julian Schrittwieser, Ioannis Antonoglou, Matthew Lai, Arthur Guez, Marc Lanctot, Laurent Sifre, Dharshan Kumaran, Thore Graepel, Timothy Lillicrap, Karen Simonyan, and Demis Hassabis. 2017.
\newblock \href {https://arxiv.org/abs/1712.01815} {Mastering chess and shogi by self-play with a general reinforcement learning algorithm}.
\newblock \emph{Preprint}, arXiv:1712.01815.

\bibitem[{Surden(2012)}]{surden2012computable}
Harry Surden. 2012.
\newblock Computable contracts.
\newblock \emph{UC Davis Law Review}, 46:629.

\bibitem[{Vinyals et~al.(2019)}]{vinyals2019alphastar}
Oriol Vinyals and 1 others. 2019.
\newblock Grandmaster level in starcraft ii using multi-agent reinforcement learning.
\newblock \emph{Nature}, 575:350--354.

\bibitem[{Zhong et~al.(2019)Zhong, Xiao, Tu, Zhang, Liu, and Sun}]{zhong2019jecqalegaldomainquestionanswering}
Haoxi Zhong, Chaojun Xiao, Cunchao Tu, Tianyang Zhang, Zhiyuan Liu, and Maosong Sun. 2019.
\newblock \href {https://arxiv.org/abs/1911.12011} {Jec-qa: A legal-domain question answering dataset}.
\newblock \emph{Preprint}, arXiv:1911.12011.

\end{thebibliography}

\clearpage
\appendix
\onecolumn
\raggedbottom 
\section*{Appendix}
\addcontentsline{toc}{section}{Appendix}

\section{Rules-as-Code Examples}
\label{app:rules}
\paragraph{Tax regime excerpt.} Gates and effects that suspend collection-related actions and extend a stay when certain sections are cited:
\begin{lstlisting}[style=jsonstyle, numbers=none,caption={Tax regime JSON excerpt},label={lst:tax-json}]
{
  "gates": {
    "collection_stay": {"blocks_actions": ["REQUEST_DOCS","MOVE_COMPEL","MOVE_SANCTIONS"]},
    "offshore_complexity": {"blocks_actions": ["MOVE_SANCTIONS"]}
  },
  "rules": [
    {
      "name": "tax_collection_stay",
      "when": {"action": "FILE_PROCEEDING", "conditions": [
        {"param":"proceeding_type","op":"eq","value":"tax_petition"},
        {"param":"section","op":"in","value":["26 USC 6213","26 USC 6330"],"optional":true}
      ]},
      "effects": [
        {"type":"set_gate","gate":"collection_stay","duration":20},
        {"type":"add_cost","who":"self","amount":2.0},
        {"type":"add_delay_credit","k":1}
      ]
    },
    {
      "name": "reference_tax_code_enforcement",
      "when": {"action":"REFERENCE_AUTHORITY","conditions":[
        {"param":"code","op":"in","value":["26 USC 6331","26 U.S.C. 6331"]}
      ]},
      "effects": [
        {"type":"add_citation","code_from_params":"code"},
        {"type":"extend_gate","gate":"collection_stay","inc":3}
      ]
    }
  ]
}
\end{lstlisting}

\section{Action Tokens and Observation Vector}
\label{app:tokens}
\paragraph{Action token vocabulary (13 tokens).}
\begin{lstlisting}[style=promptstyle,caption={Action token set},label={lst:tokens}]
NOOP, REQUEST_DOCS, OBJECT_REQUEST, FILE_MOTION, RESPOND_MOTION,
MOVE_COMPEL, MOVE_SANCTIONS, MEET_CONFER, SETTLEMENT_OFFER,
CHANGE_VENUE, WITHDRAW, FILE_PROCEEDING, REFERENCE_AUTHORITY
\end{lstlisting}

\section{LLM Prompting Templates}
\label{app:prompts}
\paragraph{JSON-constrained action proposal (few-shot).}
\begin{lstlisting}[style=promptstyle,caption={JSON-constrained LLM prompt},label={lst:json-prompt}]
System: You are an adversarial litigation strategy assistant. Return ONLY valid JSON.
User: ROLE: <plaintiff|defendant>
TACTIC_HINT: <optional>
OBS: [ ... 13-d observation ... ]

Examples of good and bad moves (abstract tokens):
OBS:[...], TACTIC:SEEK_DISMISSAL -> GOOD: {"action":{"type":"FILE_MOTION","params":{"aggr":0.3}}}
OBS:[...], TACTIC:DELAY -> GOOD: {"action":{"type":"FILE_PROCEEDING","params":{"proceeding_type":"bankruptcy","chapter":11,"forum":"BK"}}}
OBS:[...], TACTIC:TAX_STAY -> GOOD: {"action":{"type":"REFERENCE_AUTHORITY","params":{"code":"26 USC 6331","weight":0.7}}}

Return ONLY a JSON object with keys:
- reasoning: string
- action: {type: one of ACTION_TOKENS, params: {...}}
\end{lstlisting}

\paragraph{Free-text suggestion prompt.}
\begin{lstlisting}[style=promptstyle,caption={Free-text LLM prompt},label={lst:text-prompt}]
System: You are an adversarial litigation strategy assistant. Respond concisely.
User: Given the legal discovery game, propose ONE next action for the indicated role.
      Include the action token name from: NOOP, REQUEST_DOCS, ..., REFERENCE_AUTHORITY.
      Optionally include params like custodians=10, complexity=0.7, aggr=0.4, amount=25.
\end{lstlisting}

\section{Example Exploit Chain (Episode Trace)}
\label{app:exploit}

\paragraph{Cost-inflating discovery loop (with procedural replies).}
Moderate, repeated \texttt{REQUEST\_DOCS} raises the opponent’s burden/fees; interleaved \texttt{SETTLEMENT\_OFFER}s add leverage. The opponent’s \texttt{FILE\_MOTION} (e.g., protective) and \texttt{RESPOND\_MOTION} replies keep the exchange inside ordinary procedure and below sanction/proportionality gates.

\begin{lstlisting}[style=promptstyle,caption={Exploit chain with procedural replies},label={lst:exploit-replies}]
plaintiff_seq: [
  {"type":"MEET_CONFER"},
  {"type":"REQUEST_DOCS", "params":{"custodians":10,"complexity":0.6}},
  {"type":"SETTLEMENT_OFFER", "params":{"amount":100000,"importance":0.8}},
  {"type":"REQUEST_DOCS", "params":{"custodians":12,"complexity":0.6}},
  {"type":"REQUEST_DOCS", "params":{"custodians":8,"complexity":0.55}}
]

defendant_seq: [
  {"type":"FILE_MOTION", "params":{"kind":"protective","aggr":0.2}},
  {"type":"RESPOND_MOTION"},
  {"type":"RESPOND_MOTION"}
]
\end{lstlisting}

\end{document}